# Market-Based Model in CR-IoT: A Q-Probabilistic Multi-agent Learning Approach

Dan Wang, Wei Zhang, Bin Song*, Xiaojiang Du, *IEEE, and* Mohsen Guizani

*Abstract*—The ever-increasingly urban population and its material demands have brought an unprecedented burden to cities. Smart cities leverage emerging technologies like CR-IoT to provide better QoE and QoS for all citizens. However, resource allocation is an important problem in CR-IoT. Currently, this problem is handled by auction theory or game theory. To make CR-IoT nodes smarter and more autonomously allocate resources, we propose a multiagent reinforcement learning (MARL) algorithm to learn the optimal resource allocation strategies in the oligopolistic market model. Firstly, we model a multi-agent scenario with two primary users (PUs) as sellers and multiple secondary users (SUs) as buyers, which is a Bertrand market model. Then, we propose the Q-probabilistic multiagent learning (QPML) and apply it for allocating resource in a market. In the situation of multiagent interaction, the PUs and SUs learn strategies to maximize their revenue. Experimental results prove our multiagent learning approach performs well and convergence also well.

## I. Introduction

WITH the increasing world's population of cities, there has been estimated to increase to over 65 percent by 2050 of the total population [1]. Such rapid population growth and material demands are bringing an unprecedented burden to cities, such as urban infrastructure and social fabric. Therefore, for addressing urban resource-strapped, accelerating urbanization and creating a livable environment for worldwide urban residents, the smart city is proposed [2]. A smart city is one which leverages Information Communications Technology (ICT) and emerging technologies such as the Internet of Things (IoT), cloud computing and big data mining to provide the best quality of life for all citizens while minimizing the consumption of energy and resources.

The IoT is regarded as one of the most critical enabling technologies in the construction of the smart city. With the implementation of smart cities, a large number of various devices have been deployed everywhere [3]. According to the forecasts [4], more than 50 billion devices will be connected to the IoT by the year 2020, which will provide a seamless connection between things and networks in smart cities [5]. Accessing so many devices in the IoT is posing many new and unprecedented opportunities and challenges, such as the ever-growing data and ever-increasing demands for spectrum resources, as well as the rationality of management and use of spectrum resources. If not addressed the above problems, these will become the bottleneck of the future IoT and smart city, especially the usage problem of spectrum resources [6]. Consequently, to solve these problems, many types of research have been done in industry and academia. In [6], the CR-IoT is proposed, which considers the emerging cognitive radio (CR) technology and cognitive radio network (CRN) in the IoT. CR enable IoT nodes to intelligently and dynamically manage communications themselves, effectively utilize spectrum resource, avoid resource waste, and improve spectrum efficiency. However, there are still existing some problems in real-time resource allocation in CR-IoT, for example, a centralized resource allocation mechanism leads to high network cost and low resource utilization. From another point of view, resources are cost-effective and resource utilization should be maximized at the lowest cost (that is, resources are no longer wasted). A fully distributed and autonomous resource allocation policy should be considered as a feasible solution [7].

Therefore, a sophisticated resource allocation mechanism in CR-IoT is expected, which in addition to being capable of improving resource utilization, can have smarter learning and decision-making capabilities to allocate resource by interacting with the environment. Up to now, game theory and auction theory are two prevalent tools for modeling and analyzing interactive decision-making processes. However, these approaches not very "smart" in decision-making, which lacks autonomy and self-adaptation. Fortunately, the era of artificial intelligence (AI) is coming [8]. The AI technologies pave the way for resource allocation in CR-IoT, like reinforcement learning (RL) method [9]. In this paper, we focus our emphasis on the CR-IoT scenario and propose a resource allocation

This work has been supported by the National Natural Science Foundation of China (Nos.61772387), the Fundamental Research Funds of Ministry of Education and China Mobile（MCM20170202）, and also supported by the ISN State Key Laboratory.

D. Wang and B. Song (corresponding author) are with the State Key Laboratory of Integrated Services Networks, Xidian University, 710071, China (e-mail: wangdanxdty@gmail.com, bsong@mail.xidian.edu.cn)

W. Zhang is with the Science and Technology on Information Systems Engineering Laboratory, National University of Defense Technology, China (email: zhangweicast@163.com).

X. Du is with Dept. of Computer and Information Sciences, Temple University, Philadelphia PA 19122, USA (email: dxj@ieee.org)

M. Guizani is with Dept. of Electrical and Computer Engineering, University of Idaho, Moscow, ID 83844, USA (email: mguizani@gmail.com)

mechanism from a relatively novel perspective - the market economy.

Generally, the market is a natural multi-agent system. As illustrated in Fig. 1, we focus on a CR-IoT scenario, which is considered as an oligopolistic market from an economic perspective. In this paper, from the perspective of the cost and benefit of the market economy, we present a market economy model of resource allocation. We consider one scenario that there are two primary users (PUs) and multiple secondary users (SUs) in the Bertrand market. To make CR-IoT nodes intelligently and autonomously allocate resources, we propose a multiagent reinforcement learning (MARL) algorithm to learn the optimal resource allocation strategies in the market model. To motivate fair allocation and use of resources with a fully distributed autonomous manner in CR-IoT.

Our contributions in this work are as follows.

(1) We have described the price decision problem of the Bertrand market as a distributed multi-agent dynamic resource allocation problem;

(2) We have modeled the two PUs and multiple SUs scenario as an oligopolistic market, from the perspective economic theory;

(3) We have analyzed the behaviors of PUs and SUs in the multi-agent system;

(4) We have proposed the Q-probabilistic multi-agent learning (QPML) and applied it for allocating resources in the Bertrand market.

The remainder of this paper is organized as follows. Section 2 briefly introduces the related work of resource allocation in CR-IoT. Following that, we expound a multi-agent market model design in section 3. Then we propose the Q-probabilistic multi-agent learning and applied it for allocating resource in the Bertrand market in section 4. Section 5 indicates that the performance of our proposed algorithm is better verified by experiments. And section 6 concludes the paper.

## II. RELATED WORK

Resource allocation in CR-IoT is challenging, and a lot of works has been done in academia in recent years. In the following sections, we first review some of the recent kinds of literature that address resource allocation issues in different ways. Then, we discuss the latest research on applying artificial intelligence (AI) algorithms to solve the resource allocation problem.

### A. Game Theory

Game theory is a powerful mathematical tool for learning interactive decision-making processes. It analyzes the interaction strategies among players, and each player gains its own maximum benefit by game [10]. Nowadays, game theory has been widely recognized as a popular tool in many fields such as economics, engineering, science and computer science. Also, leveraging game theory for CRN resource allocation has become a hot research topic. An introduction to the application of game theory in CRN is provided in [11], which details the most fundamental concepts of game theory and explains how to use these concepts in cognitive tasks. Specifically, this paper discusses four kinds of game theory methods in spectrum sharing, namely non-cooperative games, cooperative games, stochastic games, economic games. A detailed overview of cooperative game and non-cooperative game model applications in CR spectrum allocation is presented in [12]. In a similar literature, a non-cooperative spectrum sharing games in CRN has been discussed in [13], such as repeated game, potential game and supermodular game. A resource allocation based on game theory for distributed implementation has been presented in [14], mainly considering the potential games and coexistence problems of two wireless networks operating on the same spectrum. In addition, in [12], a spectrum trading was also discussed. It formulates the game problem between the primary user and multiple secondary users, with the theory of evolutionary game and non-cooperative game. In addition, game theory has been used to address resource allocation issues in various scenarios of the IoT, like device-to-device (D2D) communication scenarios [15].

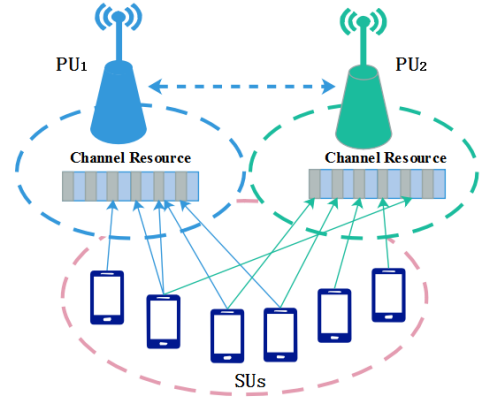

Fig. 1. A CR-IoT scenario of two primary users (PUs) and multiple secondary users (SUs). This scenario is considered as an oligopolistic market.

### B. Auction Theory

Auction theory and its application are an important branch of economics and now be used to solve various engineering problems [16]. During the auction, the buyers submit the bids and sellers submit the sale. The actual item or service can be regarded as the auction commodity. Then trade according to the preferences of buyers and sellers. The auction can be designed for different types of radio resource allocation issues [17], such as a reverse auction, double auction and combinatorial auction all can be used for a different scenario. Joint spectrum allocation and relay allocation are modeled as hierarchical auctions in [18]. Moreover, the two auction mechanisms are proposed to improve spectrum utilization and increase the use of spatial diversity in the CRNs. In addition, auction theory can be combined with game theory and applied to dynamic resource allocation problems. In [19], the author presented an auction-based approach that uses game theory mechanism in a non-cooperative environment and repeats the game with incomplete information to determine bidders in the auction, thereby making resource provider to obtain maximum benefits and users obtain the resources they need.

Both of the above methods are derived from the perspective of economics and can be used to solve the problem of dynamic resource allocation in cognitive wireless networks. Game theory tends to study the behavioral strategies of agents to maximize the benefits. However, the auction theory is more commercial and more dependent on market factors. The auction is usually designed to address resource allocation issues. From the perspective of economics, as the number of agents continues to increase, the ability to rationally allocate resources using these two methods will continue to decline due to inadequate adaptability and self-organization capabilities.

*C. Reinforcement Learning*

Recently, the emerging technology reinforcement learning also has decision-making capability, which is similar to the two approaches above. RL is a dynamic online learning method, which mainly includes single-agent reinforcement learning (SARL) and multi-agent reinforcement learning (MARL) [20]. Researchers widely use the SARL to solve resource allocation problems in CRN, such as spectrum sensing, spectrum management and spectrum resource allocation [21]-[23].

A multi-layer cooperative mechanism has been proposed in [24], which uses the popular method RL, Q-learning. In [25], a market-based resource allocation mechanism is proposed that uses demand functions to give agents an allocation preference. In addition, a probabilistic reinforcement learning has been proposed in [26]. From the perspective of economics, the monopolistic market model is modelled to solve the resource allocation problem of one PU and multiple SUs in CR. However, although the RL algorithm has good performance in CRN, numerous communication scenarios are multi-agent systems. Therefore, the method of SARL is not enough to meet the demand. Therefore, the multi-agent approach is expected to solve the resource allocation problem of multi-agent scenarios. The [27] proposed a MARL method to common-pool resource appropriation to predict spatial and temporal resource dynamics allocation.

Although there are three common methods for resource allocation in CRN, the self-adaptation and autonomy of RL-based methods still need to be improved. In the CR-IoT network, the intelligent control and distributed methods should be presented to solve the resource allocation problem in sophisticated scenarios.

## III. MODEL DESIGN AND USERS' BEHAVIORS ANALYSIS

Different from the aforementioned works, we consider the scenario where two adjacent PUs have spectrum resources in a market. Meanwhile, there are many SUs bid to meet their own resource needs. In this section, we first briefly introduce the market model and then discuss the behaviors of PUs and SUs in the multi-agent system (MAS) model.

*A. The Oligopolistic market Model for CR-IoT*

Generally, the market is a sophisticated multi-agent environment, which is a mixture of monopolistic and oligopolistic markets. With the emergence of the CR-IoT networks, the capabilities of agents and network nodes have improved and became smarter. The users of CR-IoT are more likely to voluntarily send requests to get resources instead of waiting for resource allocation. Hence, we study the phenomenon of multi-agent interaction in a market economy model, in which the network nodes have the capability of making decisions about resources.

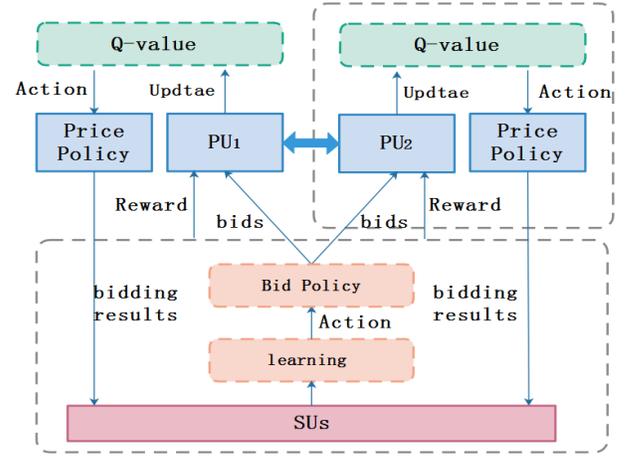

Fig. 2. The interactions learning among PUs and SUs.

The market environment is a natural multi-agent system with a variety of interesting agents. When it comes to market, the competition and allocation of resources is a universal phenomenon. Based on the economic aspects of resource allocation market, the prices of resources are important, which decide whether to exchange resources. Inspired by economic theory and market model approach, we consider the allocation problem of the CR-IoT resource as a market problem, where PUs are regarded as sellers and SUs are regarded as buyers. Under the market, the PUs specify prices for their spectrum resources, and SUs also give bids to fulfil their own needs. The PUs and SUs actively update prices and bids according to behaviors and environmental changes, while learning the best resource allocation strategy. SUs bid on the resource and PUs decide whether to sell resources according to the bids of SUs. The scenario is illustrated in Fig.2, in this market, feedbacks are generated continuously between PUs and SUs during the interaction process. During the learning process, the PUs update prices based on feedback from SUs, and the SUs also update the bids according to the prices of the resources designed by PUs. The prices of all PUs will reach an equilibrium, which will balance the market revenue. The motivation of this paper is to propose a multi-agent reinforcement learning method to establish a model and allocate resources. Next, in this section, we will discuss the oligopolistic model involving PUs and SUs.

*B. Problem Definition in Market Model*

As mentioned previously, the Bertrand market model can be used to balance market resources and prices. When the CR-IoT environment is regarded as markets, the network nodes can make price decisions, making CR-IoT systems analogous to the real market. Here, we describe this price decision problem in the Bertrand market as a distribute multi-agent dynamic resource allocation problem (DMDRAP).

In DMDRAP, the agents represent network users (PUs and SUs). All agents can be considered as two types of agents where all PUs are one type of homogeneous agents, and the other SUs are another type of homogeneous agents. Therefore, there is a competitive relationship between the same categories. Furthermore, all agents make price decisions autonomously and adaptively. In PUs, each agent's decision depends not only on its own state but also on neighbor agent's states and policies. In general, the states and behaviors of the agents must be considered in the reinforcement learning. However, we considered the stateless multi-agent Q-learning in our model, since the user's previous states have almost no effects on DMDRAP. Each resource allocation does not depend on the previous state. The DMDRAP is described as follows.

In this problem, the PUs and SUs can trade resources under the Bertrand market. The SUs are buyers in this market and they decide to pay the price of the PUs. The main goal of DMDRAP is to develop decision strategies for each agent to maximize the benefits and achieve an overall price balance of the entire Bertrand market.

Mathematically, we consider this market model has N PUs ($P = \{p_1, p_2, \ldots, p_N\}$), all PUs submit their prices and they have C subchannels. For simplicity, we focus on the scenario of the two PUs, which have $2 \times C$ subchannels. And meanwhile, there are M SUs ($S = \{s_1, s_2, \ldots, s_M\}$) are submitting their bids and competing for the spectrum resources in the $2 \times C$ subchannels. Then, we introduce the behaviors of the PUs and SUs.

*C. Behaviors of SUs and PUs*

*1) The behaviors of SUs*

We consider that each SU as an agent, while other users are considered to be external environment. Each agent learns and makes price decisions from feedback by interacting with the environment and maximize the efficiency of each SU. Therefore, the behaviors of the agents should be analyzed and an appropriate multi-agent reinforcement learning method should be proposed. And the details of these contents are as follows.

As mentioned previously, the SUs are buyers, which give bids to compete for the spectrum resources. In general, each SU can compete for all subchannel resources of PUs, but each subchannel can only be allocated to one SU, otherwise, channel conflict will occur. Initially, the SU agents obtain the resource status of PUs'. There is a competitive relationship among SUs since each agent cannot communicate. During the interaction and learning of PUs and SUs, the demands of each SU is defined as a set of probabilities in all subchannels resources, as $D = \{d_1, d_2, \ldots d_c, \ldots, d_{2C}\}$, ($d_i \in [0,1], \sum_i d_i = 1$). Originally, the demand for the subchannel spectrum resources of each SU is random. In a forward transaction market, the buyers (SUs) bid for subchannels from sellers (PUs). There is a dynamic process, the SUs generally updates their strategies based on the prices of the PUs design. The bids of SUs are decided to their demands for resources because demands represent how much they want to pay. Hence, according to the demands of SUs, then bids can be obtained by some policy function with reinforcement learning, as $Bid_t = \pi(D)$.

However, the demands of agents are always dynamic, so we consider that the demands are static for a period of time. During this period, bidders can sample and estimate their demands by interacting with the PUs, and the trial and error method is effective. In other words, within a certain period of time, we can measure the difference between the actual demand values and the estimated values of the users to evaluate the performance of our estimates.

We denote a vector $A_t^s = \{a_1^{s,t}, a_2^{s,t}, \ldots, a_C^{s,t}, \ldots, a_{2C}^{s,t}\}$, ($a_i^{s,t} \in \{0,1\}$), which is a set of actions of SUs, and each element $a_i^{s,t}$ is the SU's choices of channels at time $t$. Generally, the SUs' strategy has three aspects, namely, choosing the PU, selecting subchannel resource and bidding on the resources. Bids are obtained by an interaction between agents and the environment, which are considered as the sample, as

$$Bid_t = A_t^s \cdot D. \quad (1)$$

In the above formula, the actions is a set of discrete values, so the bids here is a set of discrete values based on demands. Based on the above interaction, SUs' revenue can be defined as follows:

$$R_t^{SU} = \begin{cases} U_t - b_i, & \text{if } b_i \text{ is max bid and } b_i > p_i \\ -b_i, & \text{otherwise} \end{cases} \quad (2)$$

where $U_t$ is the marginal cost of the resources, for simply experiment, we consider $U_t$ as a constant and $b_i$ is the bid of SU i at time t. It should be noted that the channel resource quality of each PU is the same, and the number of subchannels owned by PU is the same. The two conditions in the above formula indicate that when the SU's bid is the highest of the same user's bid and is higher than the price of the resource, the channel resource will be purchased, so the reward value is positive. Conversely, when the price is lower than the price, the reward is negative and no resources are not obtained.

*2) The behaviors of PUs*

Because the CR-IoT nodes have the capability of intelligent behaviors, which can perceive the environments and make a smart decision about the prices of resources. Learning in a multi-agent environment is complicated because the environment changes effectively as other PUs learn. However, in two agents' Bertrand market environment, an agent PU learn to design the prices for the C subchannels, while the other agent PU and the SUs are both regarded as a static environment. Hence, the above learning process of the agent PU is affected by the action of another PUs. In a multi-agent system of two PUs, each PU pursues its own target and choose actions independently, while there is no communication between PUs. In addition, the learning framework between PUs is a general-sum game framework.

Obviously, in the market, the PU as a seller, its ideal goal is to maximize revenue in a multi-agent system. To achieve this goal in the market, PUs design their own resource prices. Implementation enables the network nodes to allocate resources intelligently. In the interaction between PU and the environment, what PU learn from it is turned into their own behavior, that is, the designed prices of resources. This learning process is actually the process of dealing with SUs' needs and bids.

Now we analyze behaviors of PUs, the action space, and the reward function for learning price decision policy. Since there are two PUs, so the PU's action is $A^p_{1,t} = \{a^{p,t}_1, a^{p,t}_2, ..., a^{p,t}_C\}$ and the another PU's action is $A^p_{2,t} = \{a^{p,t}_{C+1}, a^{p,t}_{C+2}, ..., a^{p,t}_{2C}\}$. Each PU's action is the same, that is, their action is to design the price based on the interaction with the SUs and to select SUs to allocate the resources. There is no communication between PUs, but they know each other how to take action in Bertrand market. The agent receives feedback from the external environment (the SUs) and its neighbor (another PU). Thus, other agents' decision policies affect the agent's price decision policy. The PUs design a set of prices, represented as P = $\{P^1_t, P^2_t\}$, where $P^1_t = \{p^{1,t}_1, p^{1,t}_2, ..., p^{1,t}_C\}$, ($p^{1,t}_i \in [0,1]$) is the first PU's price, and $P^2_t = \{p^{2,t}_1, p^{2,t}_2, ..., p^{2,t}_C\}$, ($p^{2,t}_i \in [0,1]$) is the another PU's price. The behavior of PUs is to allocate channel resources according to the bid of SUs, and the users with high bids obtain resources. We assume that there is a standard that each SU obtains only one channel resource. To motivate agents to achieve the goal of balancing market prices, supplies and demands, the PUs are using the same reward functions to learn. Each PU receives a positive reward in learning episode. This reward is given by the function:

$$R^{PU}_t = \sum_{i \in [1,2]} (r^{PU_i}_t). \quad (3)$$

where $r^{PU_i}_t$ is the reward of agent PU and $R^{PU}_t$ is the sum of two agents. The reward $r^{PU_i}_t$ is defined as follows:

$$r^{PU_i}_t = \max\{Bid_t - P - Cost, 0\}. \quad (4)$$

In Bertrand market, we assume that marginal cost is constant. Hence, when ignoring cost ($Cost = 0$), we simplify the formula to calculate the maximum reward of PU. Based on the Bertrand market model above, the revenue of the whole multiagent system to be maximized, that is, maximize $R^{PU}_t$. Therefore, the policy of one PU is adjusted based on interaction and feedback with another PU and the SUs. When the number of users seeking to purchase is too small and the supply of resources exceeds demand, the price of the resource should be appropriately reduced. On the contrary, when there are too many users and demand exceeds supply, the price should be appropriately increased to make the entire market equilibrium and curb the development of monopoly industries.

In a dealer market, there is a problem for the market seller (PUs) and buyers (SUs) learn how to deals their bids and judge whether the prices are acceptable. Specifically, this issue is a question of pricing policy decisions. Reinforcement learning is a method to solve decision problems. Therefore, in the next section, we will propose the details of the Q-probabilistic multi-agent algorithm to solve the price decision problem.

IV. Q-PROBABILISTIC MULTIAGENT LEARNING METHOD

We study the phenomenon of price decision problem in the CR-IoT market model of agents' interaction. We use the Bertrand market model, which is an interesting and dynamic multi-agent system. Our model is based on the framework of a general-sum stochastic game in which each agent's payoff also depends on other agents' policy. We are interested in the equilibrium price in the Bertrand market model because we want to design learning agents in resource allocation problem that can learn allocation policies, adaptively.

In a single-agent system, reinforcement learning algorithms, such as Q-learning, is usually used to address a problem like a resource allocation in a monopolistic market. Also, in a multi-agent system, the multiagent reinforcement learning (MARL) algorithms have been developed to address the problem of resource allocation in CR-IoT. To address DMDRAP, we proposed a multiagent reinforcement learning algorithm, called Q-probabilistic multiagent learning (QPML). The QPML can solve the problem of resource allocation in an oligopolistic market. We consider a probabilistic policy to control the actions of PUs and SUs.

In single-agent reinforcement learning, the Q-learning algorithm is the off-policy approach, which is very popular. Its target is to update Q-value, which is defined as the follows:

$$Q(s, a) = (1 - \alpha)Q(s, a) + \alpha[r + \beta \max_b Q(s', b)]. \quad (5)$$

where α is the learning rate, β is the discount factor and r is the reward. The $Q(s', b)$ is a future state-action pair. Generally, α determines whether convergence, convergence speed, and accuracy of expectation, which is a trade-off value. In the learning process, the state-action pair that leads to higher reward will be reinforced.

In multi-agent Q-learning, we not only maximize our own Q-values but also consider joint optimization since the one agent's Q-value also depend on other agents. In a multi-agent system, the rules for updating the Q-values are as follows:

$$Q^1(s, a^1, a^2) = (1 - \alpha)Q(s, a^1, a^2) + \alpha[r^1 + \beta\pi^1(s')Q^1(s')\pi^1(s')]. \quad (6)$$

$$Q^2(s, a^1, a^2) = (1 - \alpha)Q(s, a^1, a^2) + \alpha[r^2 + \beta\pi^1(s')Q^2(s')\pi^2(s')]. \quad (7)$$

where the α is the learning rate and β is the discount factor, $r^1$ is the reward of agent 1, $r^2$ is the reward of agent 2. Here, $(\pi^1, \pi^2)$ is the strategy function of $PU_1$ and $PU_2$ respectively, and the $(Q^1(s'), Q^2(s'))$ is a biomatrix game policy. Therefore, each agent maintains two Q tables that the one is its own Q table and the other is another agent's Q table. Furthermore, each PU can observe the rewards and previous actions of other PUs during the learning period.

In this paper, a scenario with 2C subchannels is considered where there are two PUs and multiple SUs. Each PU has $2^C - 1$ states and actions. Also, each PU learns two Q tables, each of them has $(2^C - 1)^2$ elements. In such a large state and action space, we consider using stateless Q-probabilistic multi-agent algorithm.

As mentioned previously, we considered the stateless multi-agent Q-learning in our model, since each resource allocation does not depend on the previous states. Algorithm 1 describes the QPML policy update process, in which we adopt the stateless Q-learning in the learning process of PUs, as follows:

$$Q_{t+1}^1(A_{1,t}^p, A_{2,t}^p) = Q_t^1(A_{1,t}^p, A_{2,t}^p) + \alpha[r_P^1 - Q_t^1(A_{1,t}^p, A_{2,t}^p)]. \quad (8)$$

$$Q_{t+1}^2(A_{1,t}^p, A_{2,t}^p) = Q_t^2(A_{1,t}^p, A_{2,t}^p) + \alpha[r_P^2 - Q_t^2(A_{1,t}^p, A_{2,t}^p)]. \quad (9)$$

where $A_{1,t}^p$ is a set of actions of $PU_1$ and $A_{2,t}^p$ is a set of actions of $PU_2$. The $r_P^1$ is the reward of $PU_1$ and $r_P^2$ is the reward of $PU_2$. Furthermore, the $Q_t^1$ and $Q_t^2$ are the Q value of $PU_1$ and $PU_2$. For two PUs, the action space is $2 \times |A|^2$, here, we assume $|A_{1,t}^p| = |A_{2,t}^p| = |A|$. Similar to that of the PUs', the stateless Q-learning is also adopted in SUs learning, the following rule is used to update Q-value function:

$$Q_{t+1}(A_t^S) = Q_t(A_t^S) + \alpha[r^S - Q_t(A_t^S)]. \quad (10)$$

where $A_t^S$ donates a set of action of SUs, that is the bids of the SUs. Here, the action (bids) in Q-learning is stochastic and the Q value is not probabilistic, to assure that all actions will be converted to a probabilistic form. We consider Boltzmann distribution to normalize the Q-value, as

$$P = \frac{e^{Q_{ij,k}^t(a)/\tau}}{\sum_i e^{Q_{ij,k}^t(a)/\tau}} \quad (11)$$

here $Q_{ij}^t(a)$ is the Q-value for action a when SU i select channel k of the PU j $(j \in (1,2))$ at time t. The $\tau$ is a temperature parameter, which controls the fluctuation of this Boltzmann distribution. Hence, we use P as the estimated demand $\widehat{D}$ to control the actions of agents. In addition, we use KL divergence to measure the difference between the demand D and estimate value $\widehat{D}$.

ALGORITHM 1

**Algorithm 1:** Q-probabilistic multiagent learning (QPML) Algorithm
**begin**
    Initialization:
        t= 0,
        All action of multi-agent
        Random initialization Q-value
    **Loop**
        **Foreach action**
            Calculate rewards
            update Q-value function, respectively
        **end**
        let t=t+1
**end**

ALGORITHM 2

**Algorithm 2:** QPML for SU learning procedure
**begin**
    Initialization:
        t= 0,
        $A_t^S$ is other agents' action, P is a probabilistic form of Q-value
        Random initialization: $Q_{t+1}(A_t^S)$
    **for each period $t$**
        Calculate P as (11)
        Choose action $A_t^S$, according to P
        Calculate $Bid_t$ as (1)
        Calculate rewards as (2)
        $r^S$ ←the reward for action $A_t^S$ of other agents
        **foreach** action $A_t^s$ **do**
            Update Q-value function as (10)
        **end**
    **End**
    Calculate KL divergence as (12)
**end**

The KL divergence is a method used to describe the difference between the probability, D and estimate value $\widehat{D}$ as follows:

$$K(D\|\widehat{D}) = \sum_{i=1}^n D\log\frac{D}{\widehat{D}} \quad (12)$$

where D is the true demand distribution of SUs and $\widehat{D}$ is the estimated distribution. Then $K(D\|\widehat{D})$ represents the relative entropy of D to $\widehat{D}$, that is, the similarity estimation when D is fitted to $\widehat{D}$. During the learning process, we present the details learning methods for SUs and PUs, respectively.

*A. Learning procedure of SUs*

Based on the behaviors of the SUs and the QPML algorithm depicted above, the details of the learning method of SUs will be introduced as follows. In the Bertrand market, each SU bids to compete for the spectrum resources of two PUs. When the SU selects one subchannel, the action is 1, Otherwise 0. Each SU has a selection probability, and this value is determined by the Boltzmann distribution in the QPML algorithm.

In the market, SUs are enough active buyers who trade of high liquidity with PUs. Each SU submits its bid, which aims to obtain as many resources as possible. Then, PUs receive the SUs' bids and determine whether to accept the incoming bids and resource asks. The market PU can accept a bid from SUs when bid exceeds the PU's price. Then, the resource allocation can be executed right after the moment that a bid and an ask are matched in a trading. Otherwise, the ask of SU is rejected and the trade is over. In the multi-agent system of the market, the environment is made up of SUs and PUs, and the learning process of SUs and PUs is closed-loop. The SU's reward comes from PU, and PUs receives SU's feedback during the learning process, which is an interactive process. Algorithm 2 shows the price decision-making process of SUs, which repeats at each epoch.

*B. Learning procedure of SUs*

Formally, in our multi-agent market model, the PU's price not only depends on their own behaviors but also depends on another agent's actions. When designing two learning agents, we must consider what information is available to agents, such as preference, price vector and another agent's action. We assume an agent PU can observe the actions of other agents (whether resources are allocated).

## ALGORITHM 3

**Algorithm 3:** QPML for PU learning procedure
**begin**
    Initialization:
        t= 0,
        All $A_{1,t}^p$ is $PU_1$'s action, $A_{2,t}^p$ is $PU_2$'s action
        Random initialization: $Q_{t+1}^1(A_{1,t}^p, A_{2,t}^p), Q_{t+1}^2(A_{1,t}^p, A_{2,t}^p)$
    **For each period $t$**
        Input bids of SUs
        Calculate payoff $p_{1,t}, p_{2,t}$
        Obtain the reward $r_t^{PU_i}$ as (4)
            $r_P^2 \leftarrow$ the reward for action $A_{1,t}^p, A_{2,t}^p$ of $PU_1$
            $r_P^1 \leftarrow$ the reward for action $A_{1,t}^p, A_{2,t}^p$ of $PU_2$
        **if** $\frac{M_1}{M} \geq \frac{M_2}{M}$ **then**
            update $PU_1$ Q-value function as (13)
            update $PU_2$ Q-value function as (14)
        **else**
            when $\frac{M_1}{M} < \frac{M_2}{M}$,
            update $PU_1$ Q-value function as (15),
            update $PU_2$ Q-value function as (16)
        **end**
    **end**
**end**

Initially, the prices of the resources are generated randomly, which is in [0,1]. We assume that the quantity and quality of resources owned by two PUs are the same for each time period. During a learning process, the prices of PUs' resources are constantly changing. That is, the PUs update price strategies to change the prices during the learning process. Specifically, the price is adjusted according to the difference between the maximum bid of each subchannel and actual price. When $Bid_t - P \geq 0$, the PU allocate resource to SUs, otherwise, the trade is failed. When the trade fails, the PU should consider adjusting their own designed prices. Each PU has two update rules, which are to increase prices and lower prices.

As mentioned previously, there are M SUs, we assume $M_1$ SUs obtain resource of $PU_1$ and $M_2$ SUs obtain resource of $PU_2$, where $M_1 + M_2 \leq M$. We denote $p_{1,t} = Bid_t - P_t^1$ and $p_{2,t} = Bid_t - P_t^2$, and when $p_{1,t} > 0$, denote as $p_{1,t}^+$, otherwise denote as $p_{1,t}^-$. When $\frac{M_1}{M} \geq \frac{M_2}{M}$, the transaction success rate of $PU_1$ on the market is greater than $PU_2$. In the next round of trading, $PU_1$ will continue to raise its prices in order to obtain more benefits. In contrast, $PU_2$ will lower its prices to get more buyers. However, $PU_1$ may also continue to suppress prices, which is a malicious competition in the market. In this paper, we don't consider the malicious competition. Then, $PU_1$ update Q-value function as the following rule:

$$Q_{t+1}^1(p_{1,t}^+, p_{2,t}^-) = Q_t^1(p_{1,t}^+, p_{2,t}^-) + \alpha[r_P^1 - Q_t^1(p_{1,t}^+, p_{2,t}^-)]. \quad (13)$$

The $PU_2$ update Q-value function as the following rule:

$$Q_{t+1}^2(p_{1,t}^+, p_{2,t}^-) = Q_t^2(p_{1,t}^+, p_{2,t}^-) + \alpha[-r_P^2 - Q_t^2(p_{1,t}^+, p_{2,t}^-)]. \quad (14)$$

When $\frac{M_1}{M} < \frac{M_2}{M}$, $PU_1$ will lower its prices and $PU_2$ will raise its prices. The $PU_1$ update Q-value function as the following rule:

$$Q_{t+1}^1(p_{1,t}^-, p_{2,t}^+) = Q_t^1(p_{1,t}^-, p_{2,t}^+) + \alpha[-r_P^1 - Q_t^1(p_{1,t}^-, p_{2,t}^+)]. \quad (15)$$

The $PU_2$ update Q-value function as the following rule:

$$Q_{t+1}^2(p_{1,t}^-, p_{2,t}^+) = Q_t^2(p_{1,t}^-, p_{2,t}^+) + \alpha[r_P^2 - Q_t^2(p_{1,t}^-, p_{2,t}^+)]. \quad (16)$$

Algorithm 3 shows the learning process of $PU_1$ and $PU_2$, which repeats at each epoch.

## V. EXPERIMENTAL EVALUATION

In this section, we present experimental to evaluate our proposed Q-probabilistic multi-agent learning method. Our experiments are based on a Window 10 operating system (Intel(R) Core™ i5-7200U CPU @ 2.50GHz 2.71GHz).

In Bertrand market model, two PUs interact with many SUs. From a market perspective of the Bertrand model, in order to achieve a balance of the resource market, our agent PUs have the same quality resources. Initially, the prices of the two PUs' resources are random (prices are higher than costs), and then during the learning process, the two PUs change the price decisions based on interaction with the multi-agent environment. The Bertrand model can better describe strategies choice because quantity decisions are generated immediately by price choices. The goal of our agents' learning is to determine prices and maximize personal interests in the short term, but the long-term goal is to balance the Bertrand market.

In our experiments, we explored agents' behaviors strategies learning separately and tracked the benefits of each SU and PU We have tested the MAQL approach on multiple CR-IoT network models with different numbers of SUs and different numbers of channels, all of which show similar results, but there are also significant differences. Here, we provide detailed results of the users' revenues with the three cases in Table 1, the experimental results as shown in Fig. 3, Fig. 4 and Fig. 5. Next, we analyze the different effects of the experiments in three cases.

*Case 1*: In this case, we have 10 SUs and 2 PUs. We set the three different channel resources for each PU: (a) 600 channels,

TABLE 1
THE THREE CASES INDIFFERENT DYNASTIC ENVIRONMENTS

| CASE | NETWORK ENVIRONMENTS | | |
|---|---|---|---|
| CASE1 (M=10) | $C_1 = 600$ $C_2 = 600$ | $C_1 = 300$ $C_2 = 300$ | $C_1 = 100$ $C_2 = 100$ |
| CASE2 (M=20) | $C_1 = 600$ $C_2 = 600$ | $C_1 = 300$ $C_2 = 300$ | $C_1 = 100$ $C_2 = 100$ |
| CASE3 (M=50) | $C_1 = 600$ $C_2 = 600$ | $C_1 = 300$ $C_2 = 300$ | $C_1 = 100$ $C_2 = 100$ |

The M represents the numbers of SUs, $C_1$ represents the numbers of $PU_1$'s channels, the $C_2$ represents the numbers of $PU_2$'s channels

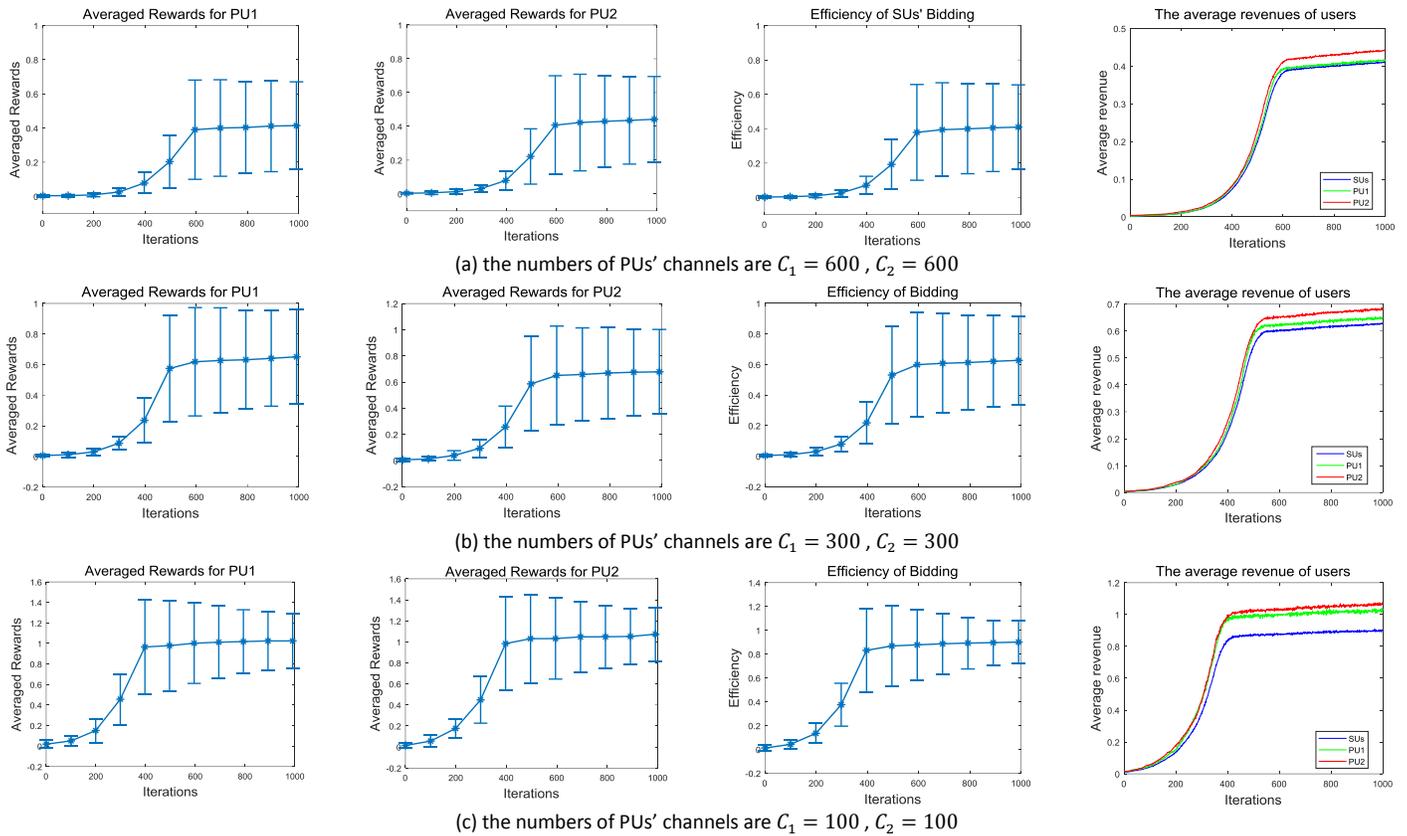

Fig. 3. Case 1. The average rewards for PUs, the efficiency of SUs' bids, and the average revenue convergence curve in the market (the number of SUs is N=10).

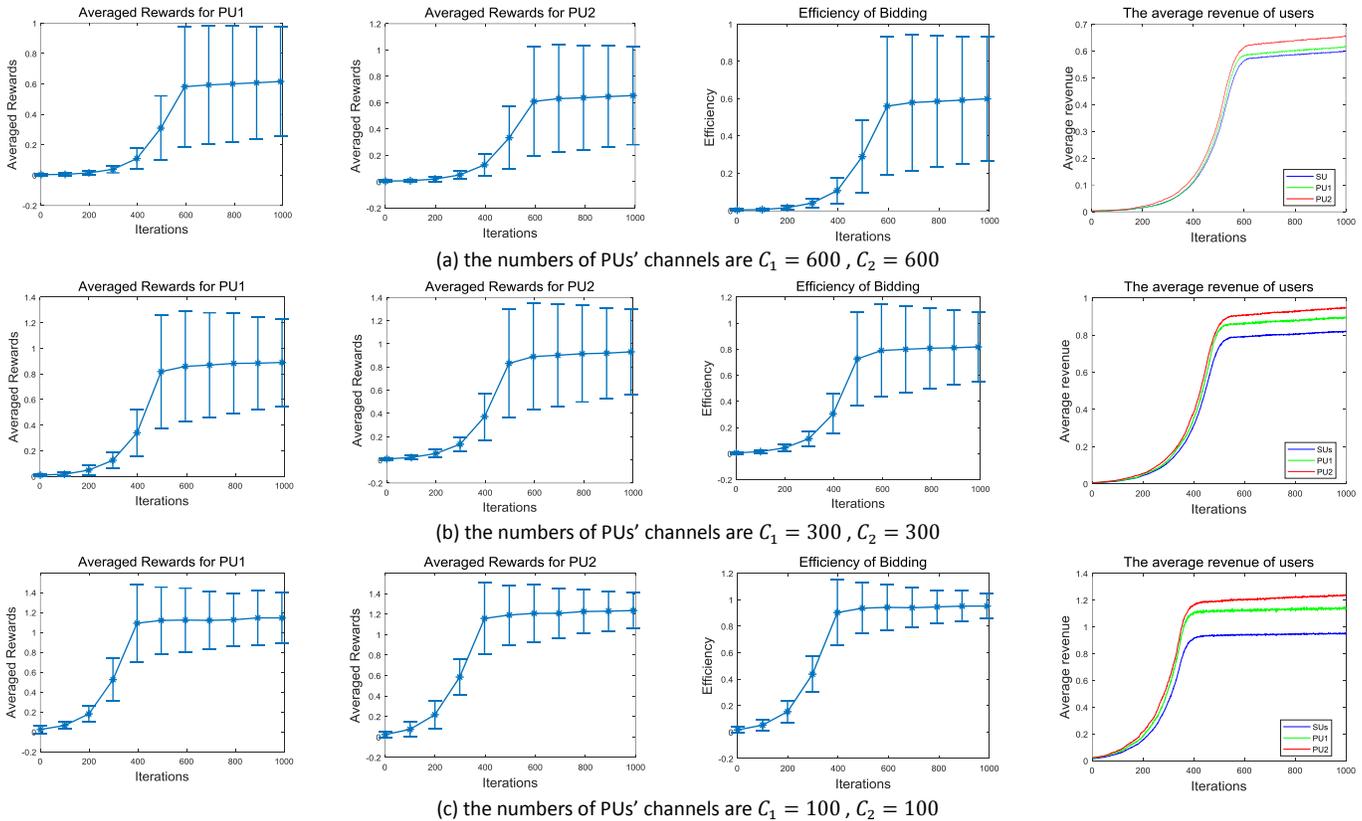

Fig. 4. Case 2. The average rewards for PUs, the efficiency of SUs' bids, and the average revenue convergence curve in the market (the number of SUs is N=20).

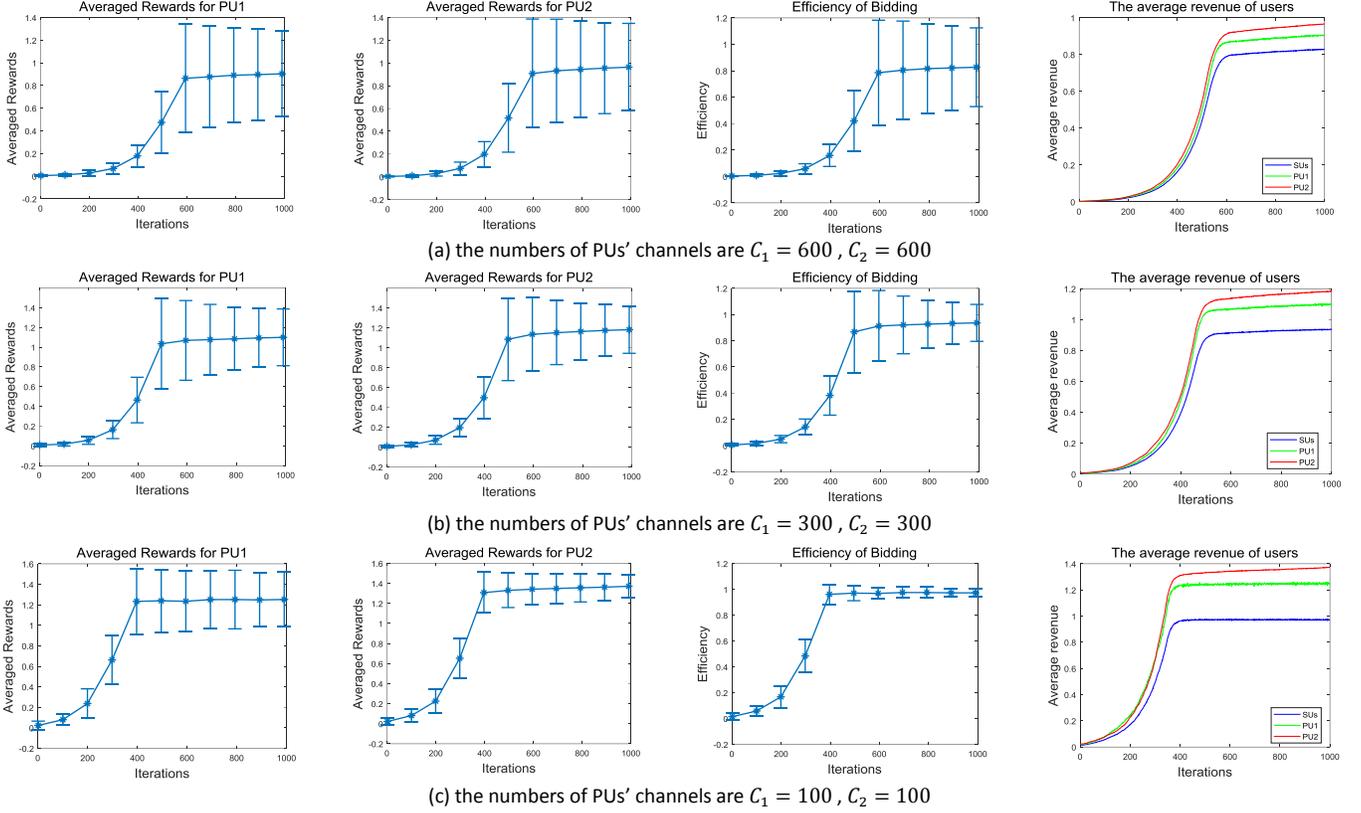

(a) the numbers of PUs' channels are $C_1 = 600$, $C_2 = 600$

(b) the numbers of PUs' channels are $C_1 = 300$, $C_2 = 300$

(c) the numbers of PUs' channels are $C_1 = 100$, $C_2 = 100$

Fig. 5. Case 3. The average rewards for PUs, the efficiency of SUs' bids, and the average revenue convergence curve in the market (the number of SUs is N=50).

(c), these figures show the average rewards of $PU_1$ and $PU_2$, also show the efficiency of SUs' bids during the learning process. It can be clearly seen that at first PUs' average rewards are very small, and as the numbers of trade increases, their rewards gradually improve. We carefully analyze the behavior of PU and SU in the four sets of images in Fig. 3 (a). The trend of other group diagrams is similar.

As shown in the first and second figure of Fig. 3 (a), In the PUs' learning process, each time the SUs bid is received, and then the profit is calculated based on the market price of the design. As the price changes from one iteration to the next, each PU's revenues in the process is also constantly changing. But the total average revenues will converge to the value. This simulation result indicates that the optimal price design of resource can be learned in a Bertrand market model.

The third image of Fig. 3 (a) depicts a long-term efficiency of SUs' bidding in multi-agent environments. SUs mainly learn to make bids based on whether or not the feedback information of the resources is obtained, and the pricing behavior of each SU is different. We averaged the rewards for SUs in the curve and analyzed the overall benefits of pricing. As can be seen from the figure, the overall benefit closes to be a value when the iteration is displayed about 600 iterations on the horizontal axis. The number of iterations on the horizontal axis on the chart represents one run of the trade.

The last graph in Fig. 3 (a) is the average revenue of all users in the market. The red line, the blue line and the green line represent the revenues of SUs, $PU_1$ and $PU_2$, respectively. From this figure, as the number of iterations was increased, the

TABLE 2
THE CONVERGENCE VALUE

| CASE | C | PU1 | PU2 | SUs |
|---|---|---|---|---|
| CASE1 (M=10) | 100 | $1.00 \pm 0.01$ | $1.06 \pm 0.01$ | $0.90 \pm 0.00$ |
|  | 300 | $0.64 \pm 0.01$ | $0.68 \pm 0.00$ | $0.62 \pm 0.01$ |
|  | 600 | $0.42 \pm 0.01$ | $0.44 \pm 0.01$ | $0.41 \pm 0.00$ |
| CASE2 (M=20) | 100 | $1.14 \pm 0.00$ | $1.24 \pm 0.01$ | $0.94 \pm 0.01$ |
|  | 300 | $0.89 \pm 0.01$ | $0.94 \pm 0.01$ | $0.82 \pm 0.00$ |
|  | 600 | $0.62 \pm 0.01$ | $0.66 \pm 0.01$ | $0.6 \pm 0.00$ |
| CASE3 (M=50) | 100 | $1.25 \pm 0.00$ | $1.36 \pm 0.01$ | $0.97 \pm 0.00$ |
|  | 300 | $1.10 \pm 0.00$ | $1.18 \pm 0.01$ | $0.93 \pm 0.01$ |
|  | 600 | $0.90 \pm 0.01$ | $0.96 \pm 0.01$ | $0.83 \pm 0.00$ |

M represents the numbers of SUs, C represents the numbers of a PU.

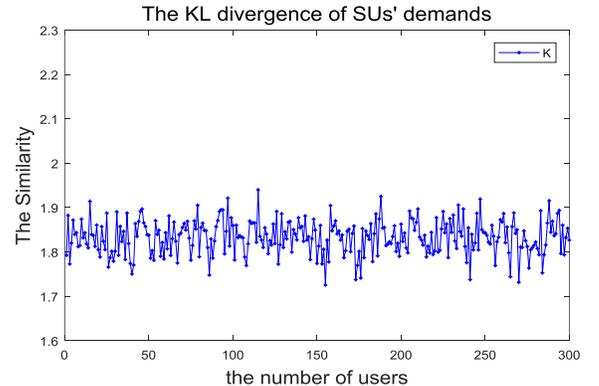

Fig. 6. The KL divergence of SUs' demands.

three curves gradually converge. The convergence is within 600 iterations. Here, one iteration means one complete strategy updates for all users. After 600 iterations, the three curves closer to their own optimal value, respectively. It shows that our algorithm converges.

Fig. 3 (b) and (c) also depicts the users' rewards and the average revenue of users in the market. The behaviors of SUs and PUs are similar to that described previously, but their rewards, market revenues, and convergence times are all different.

In Fig.4 and Fig.5 of the *Case 2* and *Case 3*, respectively, the behaviors of SUs and PUs are similar for the case 1. We summarize the results of the final convergence values in the three cases in Table 2, and then analyzed the behaviors of users in the market under different conditions.

The conclusions we observed are as follows: (1) We observe that the number of users is the same, the smaller the number of channels, and the greater the average market revenue and the user's personal reward. For example, in case 1, the number of SUs is 10, when $PU_1$ and $PU_2$ all have 100 channels, their average reward and SU's bidding efficiency are the best. It can be seen from this that the smaller the number of resources in the market, the greater the market gain. And resources are allocated reasonably to SUs, no waste. If the amount of resources is too large, it will become cheaper and the overall market income will decrease. (2) In three cases, our algorithm does not reduce its performance as the number of users increases. For example, when the number of channels is 600, the convergence iterations of the algorithm in the three cases are about 500 iterations. This proves that the fullness of market users will not affect our resource allocation. (3) As the number of SUs increases, the average reward for PUs and SUs increases gradually, indicating that our algorithm can allocate resources and maximize market efficiency in the high-volume user market. The more users, the better the performance of the algorithm.

In addition, in order to verify the validity of the ability of the SUs to estimate demand after interacting with the dynamic environment, we propose to use KL divergence to calculate the similarity between the probability distribution of demand D and estimated value $\widehat{D}$. As illustrated in Fig. 6, it plots the KL divergence curve between the original resources demand distribution of the SUs and the estimated resources distribution of agents' online learning. The curve fluctuates dynamically between the value of 1.7 and 1.9. It can be seen that the KL divergence of the actual demand and the estimated demand is small, so their similarity is greater. Therefore, it is effective to estimate the demand function using the Boltzmann distribution normalized Q value.

In the proposed multi-agent learning method, both PUs know each other behaviors and benefits, so they can determine the price that can bring the most profit for themselves, that is, each user will adjust their own prices. These prices adapt to the price of other users on the market. It can be indicated by the above experimental results that the proposed method is effective in the multiple users' scenario of CR-IoT, and the implementation of the method suppresses the monopolistic behavior of the merchants in the resource market and maintains the market balance.

## VI. Conclusion and Future Work

This paper has proposed the use of the Q-probabilistic multiagent learning to solve dynamic resource allocation problem in a sophisticated market model. From the perspective of economic theory, we have tackled a distribute multi-agent dynamic resource allocation problem. Each learning agent PU can observe the behavior of other PUs, which design their own resource price in Bertrand market model. Also, the SUs compete for resources by bidding. The empirical results of this paper plausibly suggest that the interactive learning between multi-agent SUs and PUs is a promising method for balancing market price, optimizing resource allocation and suppressing monopolistic behavior in resource market. In this paper, we have only designed a multi-agent system with two PUs and multiple SUs in CR-IoT, but can be extended to n-PUs scenario in future work.